\documentstyle[aps,epsf,psfig,twocolumn]{revtex}
\begin{document}
\draft
\flushbottom
\twocolumn[
\hsize\textwidth\columnwidth\hsize\csname @twocolumnfalse\endcsname

\title{
Kinematics of electrons near a Van Hove singularity}
\author{J. Gonz\'alez $^1$, F. Guinea $^2$ and M. A. H. Vozmediano $^3$ \\}
\address{
        $^1$Instituto de Estructura de la Materia. 
        Consejo Superior de Investigaciones Cient{\'\i}ficas. 
        Serrano 123, 28006 Madrid. Spain. \\
        $^2$Instituto de Ciencia de Materiales. 
        Consejo Superior de Investigaciones Cient{\'\i}ficas. 
        Cantoblanco. 28049 Madrid. Spain. \\
        $^3$Departamento de Matem\'aticas.
        Universidad Carlos III. 
        Avenida de la Universidad 30.
        Legan\'es. 28913 Madrid. Spain.}
\date{\today}
\maketitle
\widetext
\begin{abstract}
A two dimensional electronic system, where the Fermi surface is close
to a Van Hove singularity, shows a variety of weak coupling instabilities,
and it is a convenient model to study the interplay between
antiferromagnetism and anisotropic superconductivity.
We present a detailed analysis of the kinematics of the 
electron scattering in this model. The similitudes, and differences,
between a standard Renormalization Group approach and
previous work based on parquet summations of log$^2$ divergences
are analyzed, with emphasis on the underlying physical processes.
General properties of the phase diagram are discussed.

\end{abstract}

]
\narrowtext 
\tightenlines
Since the discovery of high-T$_{\rm c}$ superconductivity, the study of
interacting electrons in two dimensions has been a topic of
wide interest. A great variety of numerical and analytical
techniques have been developed. Progress has been relatively slow,
and there is little consensus on the phase diagram of even the
simplest models. In the following, we revisit one of the
most extensively studied models, interacting electrons in a single
band whose Fermi surface lies close to a Van Hove singularity.
Interest in this model arose rapidly after the two dimensional
character of the electron bands in the cuprates was 
established\cite{LB87,F87,S87,D87,L87}. Since then, a number of
different analytical approaches have been 
tried\cite{M89,N92,I96,G96,F98}.
Its interest increased 
when photoemission experiments showed the existence of
almost dispersionless regions near the Fermi
surface\cite{S95}, leading to renewed efforts to understand
the mechanism for superconductivity using this 
feature. 

A few facts seem to be well established about the model:
i) It is unstable, even at weak coupling, making it similar to the
one dimensional Luttinger liquids and ii) Among the possible
instabilities, antiferromagnetism and d-wave superconductivity
seem to be the most likely candidates in a wide range of
parameters, although ferromagnetism can also
exist\cite{H97,A98}. These features (apart from ferromagnetism)
where already discussed
in the early papers on the model. 
It is thus surprising the little progress made in the intervening
period.

The purpose of the present paper is to disentangle the various
factors which complicate the study of the
model, and to study the interplay and 
coexistence between superconductivity and
antiferromagnetism. As discussed in detail below, 
the apparent similarities with one dimensional systems
have lead to approximations which miss important features
of the kinematics of electrons in two dimensions. 
This neglect, in turn, gives rise to the main difficulty
in the study of the model, the fact that the perturbative
divergences mentioned before seem to go like the
square of the logarithm of the cutoff. This feature, never found in
a renormalizable field theory in the usual sense, 
has been dealt with by parquet summation techniques, using
the analogy with one dimensional systems. While these approximations
have their own merits, they cannot be considered a
proper implementation of a Renormalization Group 
procedure. In the following, we extend previous results,
and present a detailed analysis of the renormalization of
interactions in this model, using the by now standard approach
presented in\cite{S94,P94,M98}. The present work shows that
the Van Hove singularity problem can be treated by
techniques which are easily extended to generic two
dimensional problems, allowing us to extrapolate
the phase diagram to different ranges of parameters and 
fillings.

\begin{figure}
\begin{center}
\mbox{\epsfxsize 5cm \epsfbox{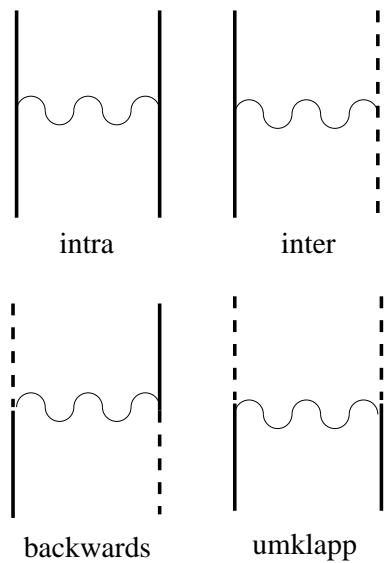}}
\end{center}
\caption{Bare couplings. Full and dashed lines stand for propagators
at the two inequivalent saddle points.}
\label{figcouplings}
\end{figure}

We focus on the weak coupling instabilities
of the model.  As mentioned earlier, a perturbation analysis
shows that, unlike models with isotropic Fermi surfaces, the
RG flow of the electron-electron interactions is not
trivial. We consider only the electronic states lying near
the two inequivalent saddle points of a square lattice.
Following preceding analyses, we classify the couplings into
four groups, shown in Fig. \ref{figcouplings}, denoted
as intrasingularity, intersingularity, backwards and Umklapp
scattering. In addition to this, the conservation of total
momentum, and the requirement that
the scattered electrons lie close to the
Fermi surface induce new constraints. The pair of outgoing momenta
is uniquely determined by the pair of incoming momenta,
except when they add up to zero\cite{S94,M98}.
This allows us
to define the Cooper pair channel in this latter instance, besides the 
forward and exchange channels, that open up when the
momentum transfer to any of the outgoing particles vanishes
(modulo $(  \pi ,  \pi )$ in the present model). 

\begin{figure}
\epsfxsize=\hsize 
\centerline{\epsfbox{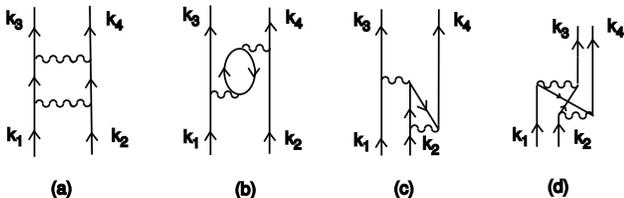}} 
\caption{Diagrams which correct the bare couplings at one loop.}
\label{fig:diagrams}
\end{figure}

The lowest order corrections to the bare couplings are shown
in Fig. \ref{fig:diagrams}. In a conventional Fermi liquid,
only the diagram (a) in Fig. \ref{fig:diagrams} depends logarithmically
on the cutoff, leading to the BCS instability. In the present case
all diagrams show logarithmic dependences, if the momentum of the 
excitations which are integrated over is close to zero. The ladder
diagram (a) in Fig. \ref{fig:diagrams} shows a $\log^2 ( \Lambda )$
dependence, while the others behave as $\log ( \Lambda )$. We are 
not considering the possibility of perfect nesting, when the Fermi surfaces
at the two singularities are parallel\cite{t-t'}.

We first discuss the diagrams (b), (c) and (d) in Fig. \ref{fig:diagrams}.
The restriction on the intermediate momenta implies that
the divergences appear when $\vec{k}_1 \approx \vec{k}_3$.
Thus, diagrams (b) and (c) give non trivial corrections in
the forward channel, and diagram (d) modifies the exchange channel.
Note however that, when all spins are parallel, the forward and exchange
channels can only be distinguished by the momentum transfer
involved, as they lead to the same final state.
Physical observables depend only on the difference between them,
like the direct and $2 k_F$ scattering between
electrons with parallel spin in one dimension.
In the following, we make use of this fact and redefine the
parallel forward channel as this combination. In this way, the exchange 
channel only has antiparallel couplings. Furthermore,
the forward channel is only renormalized by couplings in the forward
channel. 
The same applies to the exchange and Cooper pair channels. 

With the mentioned redefinition of the forward parallel couplings,
the net effect of the renormalization is given by diagram (b),
which tends to screen the bare interaction.
The RG equations in the forward channel are:

\begin{eqnarray}
\frac{\partial u_{F {\rm intra} \parallel}}{\partial l}  & = &
- \frac{1}{4\pi^2 t} c \left(u_{F {\rm intra} \perp}^2 + 
u_{F {\rm inter} \perp}^{2} + \right. \nonumber \\
&+ &\left. u_{F {\rm intra} \parallel}^2
+ u_{F {\rm inter} \parallel}^2 \right) \nonumber \\
\frac{\partial u_{F {\rm intra} \perp}}{\partial l}  & = &
-  \frac{1}{2\pi^2 t} c \left( u_{F {\rm intra} \parallel}
u_{F {\rm intra} \perp} + 
u_{F {\rm inter} \parallel} u_{{F \rm inter} \perp}
       \right)    \nonumber   \\   
\frac{\partial u_{F {\rm inter} \parallel}}{\partial l}  & = &
-  \frac{1}{2\pi^2 t} c \left( u_{F {\rm intra} \perp}  
u_{F {\rm inter} \perp}  + u_{F {\rm intra} \parallel}
u_{F {\rm inter} \parallel} \right)
\nonumber \\
\frac{\partial u_{F {\rm inter} \perp}}{\partial l}  & = &
-  \frac{1}{2\pi^2 t} c \left( u_{F{\rm intra} \parallel}  
u_{F {\rm inter} \perp} +
u_{F{\rm intra} \perp}
u_{F {\rm inter} \parallel}\right)
\nonumber \\
\frac{\partial u_{F {\rm back} \parallel}}{\partial l}  & = &
-  \frac{1}{4\pi^2 t} c' \left( u_{F {\rm back} \perp}^2 + 
u_{{F \rm umk} \perp}^{2} + \right. 
\nonumber \\ &+ &\left. u_{F {\rm back} \parallel}^2
+ u_{F {\rm umk} \parallel}^2
\right)  \nonumber\\
\frac{\partial u_{F {\rm back} \perp}}{\partial l}  & = &
-  \frac{1}{2\pi^2 t} c' \left( u_{F {\rm back} \parallel}
u_{F {\rm back} \perp}  + 
u_{F {\rm umk} \parallel} u_{F {\rm umk} \perp} 
\right)  \nonumber\\
\frac{\partial u_{F {\rm umk} \parallel}}{\partial l}  & = &
-  \frac{1}{2\pi^2 t} c' \left( 
u_{F {\rm back} \perp } u_{F {\rm umk} \perp} 
+ u_{F {\rm back} \parallel} u_{F {\rm umk} \parallel}
\right) \nonumber \\
\frac{\partial u_{F {\rm umk} \perp}}{\partial l}  & = &
-  \frac{1}{2\pi^2 t} c' \left( 
u_{F {\rm back} \parallel } u_{F {\rm umk} \perp} +
u_{F {\rm back} \perp } u_{F {\rm umk} \parallel} 
        \right)  \nonumber \\ & &     \label{flow}
\end{eqnarray}
where $c , c'$
are the prefactors of the polarizabilities at zero and 
${\bf Q} = ( \pi , \pi )$
momentum transfer. 
A formally equivalent equation is obtained
for the couplings in the exchange channel, except that, in this case,
the signs in the r. h. s. of Eq. (\ref{flow}) are positive,
and repulsive couplings increase upon scaling.
The equations in (\ref{flow}) can be further simplified.
For instance, we can add and substract the upper
four equations, to obtain:
\begin{eqnarray}
\frac{\partial \left( u_{F {\rm intra} \parallel}
\pm u_{F {\rm inter} \parallel} \right)}{\partial l}  & = &
-  \frac{1}{4\pi^2 t} c  \left[ \left( u_{{F \rm intra} \perp} \pm 
u_{{F \rm inter} \perp}
       \right)^2   + \right.  \nonumber   \\
&+ &\left. \left( u_{F {\rm intra} \parallel} \pm
u_{F {\rm inter} \parallel} \right)^2 \right] \nonumber \\   
\frac{\partial \left( u_{F {\rm intra} \perp}
\pm u_{F {\rm inter} \perp} \right)}{\partial l}  & = 
& -  \frac{1}{4\pi^2 t} c  
\left( u_{{F \rm intra} \parallel} \pm 
u_{{F \rm inter} \parallel} \right) \times \nonumber \\ & 
&\left( u_{{F \rm intra} \perp} \pm 
u_{{F \rm inter} \perp} \right) 
\label{flow2}
\end{eqnarray}
\begin{figure}
\epsfxsize=\hsize 
\centerline{\epsfbox{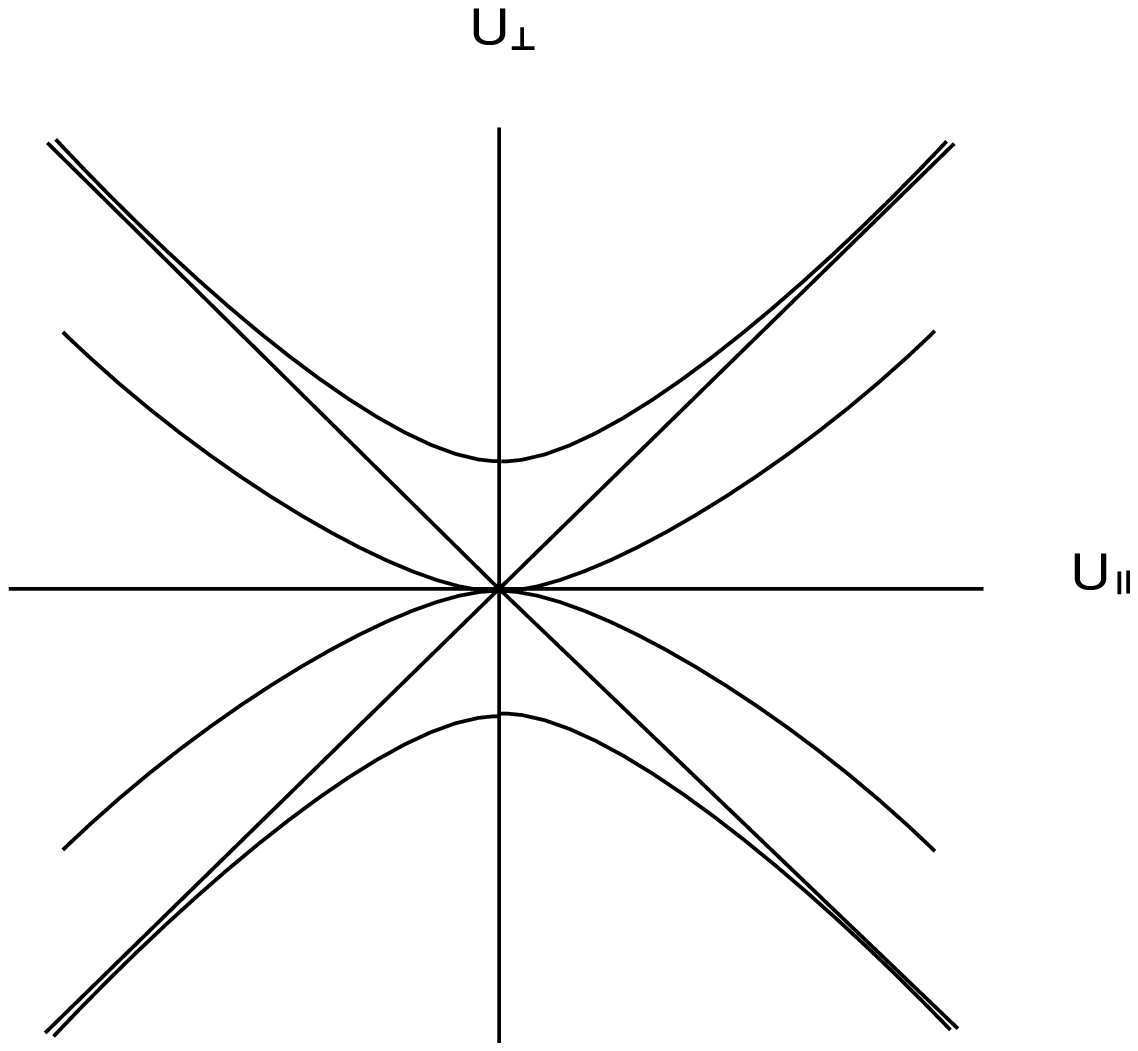}} 
\caption{Generic flow of the couplings. The flow direction is towards the
left in the forward channel, for the combination of couplings in Eq.
(\ref{flow2}). A similar
flow diagram can be drawn for the 
Cooper pair channel, with the replacement of $u_{\parallel}, u_{\perp}$ by
the Cooper pair couplings $u_{C {\rm intra} \perp}, u_{C {\rm umk} \perp}$.}
\label{fig:flow}
\end{figure}

These equations can be integrated analytically. The  main features of the
flow are depicted in Fig. \ref{fig:flow}. The flow for the
forward channel is towards negative values of the abscissas, the
opposite as for the exchange channel. The hamiltonian flows towards
different fixed points for different initial conditions, defining
different universality classes. When the initial couplings
are spin independent\cite{S87,L87,F98}, the flow starts at the
diagonal. The couplings in the forward channel vanish in the
infrared. If the bare hamiltonian is
the Hubbard model in its usual form, the initial parallel
couplings are zero. Then, the couplings must flow towards a
strong coupling regime. If only the forward and exchange channels
are considered, the leading instability is towards a broken symmetry
state with antiferro- or ferromagnetic order, depending on whether
$c'$ or $c$ is larger. It is easy to
show that the (attractive) divergent flow of the parallel   
couplings in the forward channel matches the repulsive
flow of the antiparallel couplings in the exchange channel.
The two channels renormalize the parallel and transverse 
spin-spin correlations, respectively, so that the flow maintains the original
SU(2) symmetry of the microscopic model.
In the following, we consider the
modifications to this picture induced by the inclusion
of the Cooper channel, which has to be considered independently.

The main source of difficulties,
the $\log^2$ divergences in the 
perturbation expansion, is found in the Cooper channel. A number of previous
works have dealt with them by scaling the couplings as function
of $\log^2 ( \Lambda )$, instead of the more standard $\log ( \Lambda )$.
Let us consider in some detail the implications of this
procedure. In a renormalizable theory in the usual sense,
an effective hamiltonian can be defined at each scale,
$\Lambda$, with dimensionless interactions, $\{ \tilde{u}_i \}$,
that depend on $\Lambda$. The elegance of the method resides in the fact
that there is no explicit dependence on $\Lambda$ elsewhere.
The most general equation for the flow of the
couplings is:
\begin{equation}
\Lambda \frac{\partial \tilde{u}_i}{\partial \Lambda} 
= f \left( \{ \tilde{u}_j \} , \frac{\Lambda}{\Delta} \right)
\label{scaling}
\end{equation}
where the functions $f$ are dimensionless and $\Delta$
stands for additional low energy scales.
The statement that the model is renormalizable
implies that the limit
$\Delta / \Lambda \rightarrow 0$
can be safely taken. On the other hand, a scaling with
$\log^2$ implies that the r. h. s. of Eq. (\ref{scaling})
is of the type:
\begin{equation}
f \left( \{ \tilde{u}_j \} , \frac{\Lambda}{\Delta} \right) 
= \tilde{f} ( \{ \tilde{u}_j \} )
\log \left( \frac{\Lambda}{\Delta} \right)
\end{equation}
The dependence on the low energy scales represented by
$\Delta$ cannot be avoided. A scaling in terms of $\log^2$
is not equivalent to a conventional RG calculation, as
it ignores the low energy scales present in the flow equations.
It resembles more a parquet summation of leading divergences.
In one dimension, where only logarithmic divergences appear,
both approaches are equivalent. This is not the case in the 
present problem. It can be shown\cite{G96} that a $\log^2$
divergence, when inserted into a self energy diagram,
leads to further $\log^2$ divergences in the one
electron propagator. The usual wavefunction renormalization
cannot take care of this divergence, signalling again
the existence of hidden scales and/or new, non local
interactions.

In order to avoid the $\log^2$ divergences in the Cooper
channel, we move the chemical potential slightly away
from the singularity. It is easy to show that the
divergence in the Cooper channel becomes
$\log ( \Lambda / \omega ) \log ( \Lambda / \mu )$, where
$\mu$ is the chemical potential. If $\mu \ll \Lambda$,
the other channels are unaffected. 
The factor $\log ( \Lambda / \mu )$ is simply the density
of states at the Fermi level, which, in principle, can also
be scale dependent. A full treatment of this term requires
the analysis of the flow of the chemical potential itself.

We will assume that the system is in contact with an
external reservoir which fixes the physical,
fully renormalized chemical potential.
The effective chemical potential which must be
inserted into the low energy hamiltonian at a
scale $\Lambda$ is renormalized by a Hartree diagram\cite{S94}.
In the present case, this diagram has a logarithmic singularity,
besides the usual linear dependence on $\Lambda$ dictated
by dimensional considerations\cite{S94,G96}.
As a result, one finds for the dimensionless
chemical potential, $\tilde{\mu} = \mu / \Lambda$:
\begin{equation}
        \frac{\partial \tilde{\mu}}{\partial \log ( \Lambda )} =
        \tilde{\mu} - \left( \tilde{u}_{F {\rm intra}} +
	\tilde{u}_{F {\rm inter}} \right) \log \left(
        \frac{1}{1 - \tilde{\mu}} \right)
\label{Fermi}
\end{equation}
where a summation over parallel and perpendicular spin orientations
is assumed, and the couplings have been written in a dimensionless
form.
This equation has a stable fixed point, $\tilde{\mu}^*$. This fixed
point is attained if $\tilde{\mu} > 0$, that is, if the chemical
potential lies above the singularity. The existence of this
fixed point is crucial in simplifying the treatment of
the scaling in the Cooper channel. For sufficiently
small initial couplings or at sufficiently low energies,
we can replace the $\log ( \Lambda / \mu )$ in the scaling of
the couplings in the Cooper channel by the
constant $\log ( 1 / \tilde{\mu}^* )$.
Note that the previous analysis does not contradict the conservation
of the number of electrons implied in Luttinger's theorem,
as we have chosen to work at fixed external chemical potential.
We find that this is the natural choice in the problem at hand.
The restriction of the study to small patches near the saddle
points is justified as these regions give rise to the main
divergences in perturbation theory. However, the remaining
sections of the Fermi surface are, at least, a reservoir of electrons,
always in contact with the system studied here.

The previous discussion shows 
that one of the two $\log ( \Lambda )$ factors which appear 
in the Cooper channel should not be included in the
RG equations, provided that the chemical potential is
shifted slightly away from the singularity.
The resulting equations are analogous to those
in Eq. (\ref{flow}), except that the coefficients
include a constant equal to  $\log ( 1 / \tilde{\mu}^* )$ 
at the beginning of the scaling. Note, finally, that
the existence of a finite chemical potential implies
that the flow in Eqs. (\ref{flow}) ceases to be valid
at $\omega \sim \mu$. On the other hand, the couplings reach a strong
coupling regime at scales $E_c \sim \Lambda e^{- 1 / u_0}$
where $u_0$ stands for the initial dimensionless
couplings. Hence, the phase diagram, which is determined by
the couplings which reach first the strong coupling limit,
should be independent of $\mu$, provided that $\mu \ll E_c$.

We now analyze the phase diagram by identifying the couplings which
diverge at the highest scale. A simple inspection shows that
ferromagnetism prevails when $c > c'$, associated with
the divergences in $u_{F {\rm intra} \parallel}, u_{F {\rm intra} \perp}, 
u_{F {\rm inter} \parallel}$ and $u_{F {\rm inter} \perp}$. 
When $c < c'$ the leading divergences
are either in the set 
$ \{ u_{F {\rm back} \parallel}, u_{F {\rm back} \perp},    
u_{F {\rm umk} \parallel}$, $u_{F {\rm umk} \perp}$ \}
or in the couplings for the Cooper channel
$u_{C {\rm intra} \perp}$ and $u_{C {\rm umk} \perp}$.
The first set gives rise to antiferromagnetism, while
the second pair leads to d-wave superconductivity, even if
the initial couplings are repulsive, provided that
$u_{C {\rm umk} \perp} > u_{C {\rm intra} \perp} > 0$  as shown 
in Fig. \ref{fig:flow}.
In the Hubbard model, these couplings are initially equal.
Finite corrections, not considered here, influence differently
$u_{C {\rm umk} \perp}$ and $u_{C {\rm intra} \perp}$,
leading to the conditions for d-wave superconductivity.
These finite corrections are those considered by Kohn
and Luttinger in their pioneering work on anisotropic superconductivity
from repulsive interactions\cite{KL65}.

Thus, the phases which arise in a weak coupling treatment of
the Hubbard model near a Van Hove singularity are ferromagnetism
antiferromagnetism and d-wave superconductivity.
The competition between antiferromagnetism and
superconductivity depends on the value of $c'$ and 
$\log ( \mu^* )$. As both types of couplings diverge,
the transition between the two is discontinuous
( the phase diagram in the strong coupling limit to which the
system flows should be well described within a mean field
approximation with the appropiate couplings\cite{MF98}).
Finally, the charge compressibility depends on the 
density of states at the Fermi level divided by one plus the
sum of the parallel and antiparallel couplings in
the intrasingularity forward channel. As discussed earlier, 
parallel couplings flow towards divergent negative values, while
antiparallel couplings are divergent and positive. The sum 
tends to zero. Thus, the compressibility approaches the
density of states, which, by definition, is large.
We find no evidence of a vanishing compressibility\cite{F98}.
On the other hand, we cannot rule out that the
cancellation found here does not exist when the Fermi level
is sufficiently far from the singularity, leading
to an infinite compressibility
and phase separation\cite{M99}.  

In conclusion, we have analyzed in detail the kinematics of electrons 
near a two dimensional Van Hove singularity. While the model
has been studied extensively, the classification of the
different allowed channels has never been presented. 
This framework, which makes connexion with standard applications
of the Renormalization Group program to two and three
dimensional electronic systems\cite{S94}, clarifies the
different physical processes involved, and the main features
of the phase diagram. The results can differ substantially
from those obtained by summing, in an uncontrolled
way, $\log^2 ( \Lambda )$ divergences. Finally, it can be easily extended
to address other types of anisotropic Fermi surfaces\cite{GGV97,ZS96}.


\begin{references}
\bibitem{t-t'}
If we use a tight binding scheme to describe the electron bands,
the case of parallel Fermi lines corresponds to $t' = 0$, in
the usual language. All available evidence suggests that
the model is antiferromagnetic with long range order at
zero temperature.
\end{references}
\end{document}